# Media Content Delivery Protocols Performance and Reliability Evaluation in Cellular Mobile Networks

Kirill Krinkin, Igor Dronnikov
Saint Petersburg Electrotechnical University "LETI"
Saint-Petersburg, Russia
email: kirill@krinkin.com, dronnikovigor@gmail.com

*Abstract*—Currently, tens of millions of devices around the world communicate with/ each other via cellular networks. In this paper, we study the stability of network content delivery protocols to the effects of network interference. To conduct the research, a tool was developed that allows testing of protocols, such as TCP, UDP and QUIC. The analysis and comparison of the obtained test results was carried out. In the conclusion, the best protocols for the content delivery were shown.

I. INTRODUCTION

It is difficult to imagine the modern world without means of communication. Every day, tens of millions of devices around the world communicate via computer networks. However, as the number of devices increases, so does the number of terms of use for these devices. So, some users of modern means of communication have a good stable connection, while other users may have problems connecting. An unstable network connection usually occurs when using wireless networks inside reinforced concrete buildings, or outside of large cities where the network infrastructure is not very well developed. As a result, device users may receive certain content with delays, or only partially - due to a connection failure.

At the same time, each network is organized according to the appropriate standards. All devices communicate according to the generally accepted Open Systems Interconnection model (OSI) standard, usually using the most popular transport protocols Transmission Control Protocol (TCP) or User Datagram Protocol (UDP). This is why any company that generates a large amount of traffic sooner or later thinks about what protocol it should use in its work, so that customers who would like to receive the content of this company, get it quickly and in full, regardless of the quality of connection on the client's device. Thus, the relevance of the problem lies in the complexity of choosing the appropriate protocol, since each protocol has its own set of features, in particular, different resistance to network interference. In accordance with this, it is necessary to study the stability of TCP and UDP protocols to the effects of network interference, as well as to compare them with the new protocol from Google - Quick UDP Internet Connections (QUIC) [1].

Most people [2] access social networks from mobile devices via cellular networks. The most common access pattern is the request media (image and video thumbnails) for the content feed. It motivated us to study the performance and reliability of the content delivery protocols via cellular networks (2G, 3G, 4G, Long-Term Evolution (LTE) etc).

In Section 2, a study of the current state of the problem was conducted, and works on this topic were studied. Section 3 describes network simulation using the developed tools and we provide a description of the protocols under test, including a brief description of how they work. In Section 4, we provide a description of the test cases during the experiment. Also, we give an analysis of the results obtained during the experiment.

II. RELATED WORKS

Since the problem described in the previous section has been relevant for a long period of time, there are a number of works describing the pros and cons of the TCP and UDP protocols, as well as their comparison. For example, AL-Dhief et al. [3] studied the performance of TCP and UDP protocols in order to identify the best protocol. In this work, attention is paid to such network parameters as delay, network throughput, delivery ratio of packets and packet loss ratios. The authors of this work tested the protocols on two scenarios and found that TCP is more efficient and reliable than UDP. The advantages of this work include detailed theoretical calculations, a description of the testing method and a test bench. This paper also describes the calculation of metrics and provides the resulting graphs. The disadvantages of this work include a small number of test scenarios - only changes in the speed parameters and data sizes were considered. In addition, parameters such as delay and packet loss are considered only as the results of the action of the two parameters being changed, and not as the cause.

Another paper, Coonjah et al. [4] examine the performance of TCP and UDP protocols inside TCP and UDP tunnels. The authors conclude that TCP in the UDP tunnel provides better latency. Also, in this work, a series of tests were performed, UDP traffic was sent inside the UDP tunnel and the TCP tunnel sequentially. The same tests were performed using TCP traffic. The advantages of the work include a detailed theoretical description of the protocols, as well as the test stand. This paper contains a detailed description of test scenarios and resulting graphs. Disadvantages include insufficient coverage of the impact on performance of other parameters, such as latency, packet loss, and network speed, as well as the possible presence of network interference in the test stand using physical switching between devices.

The next paper, M. P. Sarma [5] considers the issue of performance evaluation based on modeling of transport layer protocols TCP and UDP using two popular queue management methods: Random Early Detection (RED) and Tail Drop. Performance is evaluated in terms of throughput, queue latency, packet drop rate, and bandwith usage. In the conclusions, the author points out that the simulation results show that RED is superior to Tail Drop in terms of queue latency and packet drop rate. However, the performance of queue management algorithms also depends on the protocols they are applied to, TCP or UDP. The type of network topology, whether it is a shared UDP and TCP topology, or just one type of client topology also affects the performance of buffer management. For some applications, UDP using the RED queue will provide better performance, and for reliable packet delivery, TCP using the RED queue will be better than other protocols and queues in a high-speed Local Area Network (LAN). The positive aspects of this work include a detailed description of the work of queues and the test stand, as well as a detailed review of the results and the variability of test scenarios. The disadvantages of this work include that parameters such as delay and packet loss are considered only as side effects.

Another paper, S. A. Nor et al. [6] analyze the transport layer protocols that are used for video streaming. Furthermore, through simulation results, performed in Network Simulator – 3 (NS-3), the strength and weaknesses of TCP, Scalable TCP (SCTP), Datagram Congestion Control Protocol (DCCP) and UDP are presented that may give the idea of selecting the best protocols for the LTE environment. Also, this performance evaluation can provide a base to determine which protocol can be better for which metrics among the four, i.e., end to end delay, throughput, packet loss, and average jitter. The advantages of the work include good theoretical description of measurements and graphs for results. In addition, it would be interesting to see comparison with 2G, 3G, Wi-Fi, though it is not mentioned in paper state.

As seen in this section, despite the fact that the question of comparing the performance and resistance of TCP and UDP protocols to interference has been open for a long time, the results of the work still vary for different situations. Therefore, the task of studying the stability of protocols to the effects of cellular mobile networks interference is relevant.

III. EVALUATION APPROACH

The NetPacket Simulator tool [7] was developed to solve the problem of investigating the stability of protocols to network interference. This tool can be used to describe various network connection configurations and emulate possible network interference. For example, this tool can be used to see how the following parameters affect data transmission:

- RTT - round-trip time, the time it takes for the data packet to reach the recipient and the confirmation of receipt to reach the sender.
- Bandwidth - maximum data transfer speed over the network.
- Packet Loss - packet loss occurs when one or more data packets passing through the computer network do not reach their destination. Packet loss is caused by errors in data transmission, usually over wireless networks, or network congestion. Packet loss is measured as the percentage of lost packets relative to sent packets.
- Upload/Download rate - download speed is the speed at which data is transferred from the Internet to the user's computer. The upload speed is the speed of data transfer from the user's computer to the Internet. Internet companies often provide an asymmetric communication channel by default - loading is faster than return. The reason for this is that most people need to download information more. This allows the user to quickly download movies, music, and a large number of documents.
- Multiple clients - running multiple clients on the same network.

The developed tool works on the basis of virtual network drivers TUN/TAP. TUN/TAP is used to provide packet reception and transmission for user space programs. TUN (stands for network TUNnel) is a network layer device and TAP (stands for network TAP) is a link layer device. Both of them are virtual network kernel devices — this allows one to organize a virtual network within a single physical device, which eliminates the occurrence of uncontrolled network interference, if the network is organized between several physical devices.

Before conducting the experiment, it should be noted that the TCP data transfer protocol has a built-in implementation of data receipt verification and in case of loss of any packets, it will independently re-forward the lost data and guarantee their delivery. Therefore, a TCP connection can be organized directly "out of the box" - one just needs to describe the connection setup on the client and server and be sure of 100% data delivery.

In the case of the UDP data transfer protocol, the situation is different. Although a UDP connection can be set up right out of the box, UDP does not implement data receipt checks and does not guarantee that the client will receive them. Thus, if one needs to make sure that the data reached the client in its entirety or some part was lost, one must implement data integrity checks, re-forwarding, and other mechanisms oneself.

One of the add-ons for the UDP protocol with its own implementation of all mechanisms is QUIC. As shown in Figure 1, QUIC is located under HyperText Transfer Protocol 3 (HTTP/3). It partially replaces the HyperText Transfer Protocol Secure (HTTPS) and TCP layers, using UDP for packet generation. QUIC only supports secure data transfer, since Transport Layer Security (TLS) is fully embedded in QUIC.

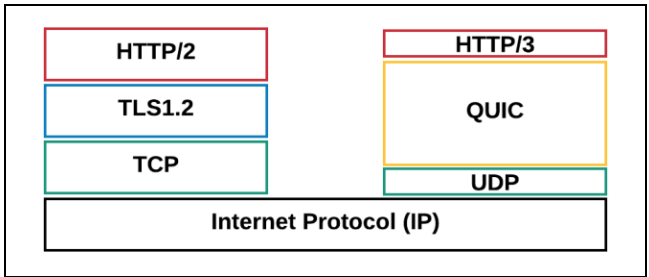

Figure 1. TCP and QUIC protocol stack

Since the purpose of our research is to compare the TCP, UDP and QUIC protocols, it is necessary to implement a full-fledged comparison for the UDP protocol, as well as mechanisms for establishing a connection, making up for losses, and confirming data receipt. This is because, for any potential use of the UDP protocol, it will be necessary to implement these mechanisms if the ultimate goal is guaranteed delivery of content to the client. Thus, our implementation of the self-made UDP (smUDP) protocol that will participate in testing will have the following mechanisms implemented: Error Correction, Pacing, Flow Control, and Congestion Control.

## IV. EVALUATION

To study the stability of protocols to network interference, the following experimental scenarios were identified. Among network interference cases we selected: RTT, Bandwidth, Packet Loss, Upload/Download rate, and Congestion Control window size. These scenarios reflect the most popular network connections. The information for the scenarios was collected by studying the statistics of usage by real social network users [8]:

A. Wi-Fi
- RTT - 110 ms
- Packet Loss - 0.5%
- Bandwidth - 2.2 Mbit/s
- Upload/Download rate - 0.7
- Congestion Control window - 1 mbyte
- File Size - 50, 100, 250, 500, 1000 Kbytes

B. LTE
- RTT - 250 ms
- Packet Loss - 0.7%
- Bandwidth - 2.0 Mbit/s
- Upload/Download rate - 0.7
- Congestion Control window - 1 mbyte
- File Size - 50, 100, 250, 500, 1000 Kbytes

C. 3G
- RTT - 550 ms
- Packet Loss - 0.5%
- Bandwidth - 1.0 Mbit/s
- Upload/Download rate - 0.7
- Congestion Control window - 1 mbyte
- File Size - 50, 100, 250, 500, 1000 Kbytes

D. 2G
- RTT - 900 ms
- Packet Loss - 2.5%
- Bandwidth - 0.2 Mbit/s
- Upload/Download rate - 0.7
- Congestion Control window - 1 mbyte
- File Size - 50, 100, 250, 500, 1000 Kbytes

E. RTT influence
- RTT - 10, 50, 100, 250, 500, 750, 1000 ms
- Packet Loss - 0.5%
- Bandwidth - 2.2 Mbit/s
- Upload/Download rate - 0.7
- Congestion Control window - 1 mbyte
- File Size - 250 Kbytes

F. Packet Loss influence
- RTT - 100 ms
- Packet Loss - 0.5, 1.0, 1.5, 2.0, 2.5%
- Bandwidth - 2.2 Mbit/s
- Upload/Download rate - 0.7
- Congestion Control window - 1 mbyte
- File Size - 250 Kbytes

G. Bandwidth influence
- RTT - 100 ms
- Packet Loss - 0.5%
- Bandwidth - 0.2, 0.6, 1.0, 1.4, 1.8, 2.2 Mbit/s
- Upload/Download rate - 0.7
- Congestion Control window - 1 mbyte
- File Size - 4096 Kbytes

Based on these scenarios, network connection configurations were described and the average time required for query execution to get the required amount of data was measured. The average time per request was calculated based on data on the execution time of five repeated simulations.

After the experiments had been performed, the results were summarized in the following Tables I - VII:

TABLE I. WI-FI

| File Size, kbytes | smUDP, ms | QUIC, ms | TCP, ms |
|---|---|---|---|
| 5 | 137 | 373 | 206 |
| 100 | 163 | 492 | 538 |
| 250 | 234 | 597 | 1 442 |
| 500 | 278 | 741 | 2 436 |
| 1000 | 563 | 1 386 | 4 784 |

TABLE II. LTE

| File Size, kbytes | smUDP, ms | QUIC, ms | TCP, ms |
|---|---|---|---|
| 5 | 304 | 594 | 447 |
| 100 | 408 | 966 | 1 310 |
| 250 | 516 | 1 370 | 3 395 |
| 500 | 530 | 2 063 | 7 182 |
| 1000 | 1 052 | 3 231 | 11 682 |

TABLE III. 3G

| File Size, kbytes | smUDP, ms | QUIC, ms | TCP, ms |
|---|---|---|---|
| 5 | 663 | 1 136 | 956 |
| 100 | 777 | 2 044 | 1 767 |
| 250 | 1 115 | 2 437 | 4 086 |
| 500 | 1 129 | 3 580 | 5 852 |
| 1000 | 2 248 | 5 383 | 12 439 |

TABLE IV. 2G

| File Size, kbytes | smUDP, ms | QUIC, ms | TCP, ms |
|---|---|---|---|
| 5 | 1 083 | 2 196 | 1 130 |
| 100 | 1 809 | 4 121 | 6 229 |
| 250 | 2 575 | 5 900 | 14 637 |
| 500 | 4 575 | 10 549 | 27 493 |
| 1000 | 8 403 | 17 379 | 59 152 |

TABLE V. RTT INFLUENCE

| RTT, ms | smUDP, ms | QUIC, ms | TCP, ms |
|---|---|---|---|
| 10 | 38 | 91 | 214 |
| 50 | 114 | 300 | 721 |
| 100 | 233 | 546 | 1 355 |
| 250 | 516 | 1 297 | 3 082 |
| 500 | 1 018 | 2 212 | 3 916 |
| 750 | 1 518 | 3 316 | 4 220 |
| 1000 | 1 816 | 4 505 | 5 411 |

TABLE VI. PACKET LOSS INFLUENCE

| PL, % | smUDP, ms | QUIC, ms | TCP, ms |
|---|---|---|---|
| 0.5 | 198 | 688 | 1 334 |
| 1 | 218 | 728 | 1 701 |
| 1.5 | 237 | 756 | 2 144 |
| 2 | 220 | 852 | 2 721 |
| 2.5 | 237 | 929 | 3 145 |

TABLE VII. BANDWIDTH INFLUENCE

| Bandwidth, Mbit/s | smUDP, ms | QUIC, ms | TCP, ms |
|---|---|---|---|
| 0.2 | 29 556 | 29 707 | 29 567 |
| 0.6 | 9 847 | 10 135 | 17 792 |
| 1 | 5 843 | 9 450 | 17 754 |
| 1.4 | 4 131 | 9 544 | 17 730 |
| 1.8 | 3 183 | 9 430 | 17 971 |
| 2.2 | 2 588 | 9 516 | 17 758 |

The obtained data on the stability of each data transfer protocol to network interference are shown in Figures 2 – 5, 7 – 9.

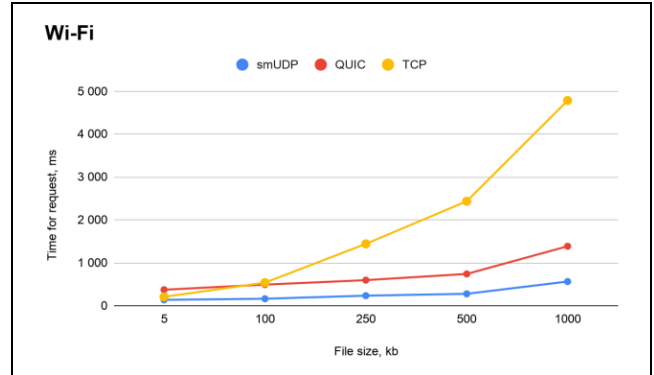

Figure 2. Wi-Fi

In Figures 2 through 5, we can see that the TCP protocol has worse behavior than other protocols. While for QUIC and smUDP, the increase in request time increases linearly with the increase in the size of the requested file, for TCP, there is an exponential increase.

TCP is designed in such a way that TCP generally uses a TCP 3-way handshake [9]: the sender sends a SYN packet, waits for a SYN-ACK packet from the recipient, then sends an ACK packet.

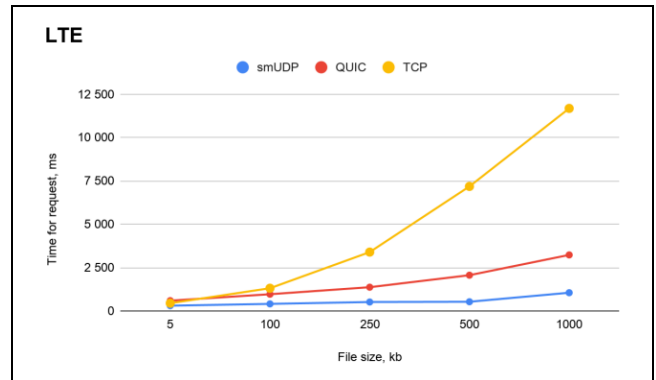

Figure 3. LTE

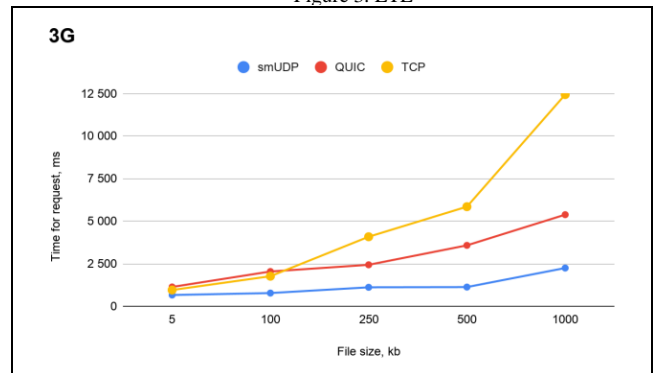

Figure 4. 3G

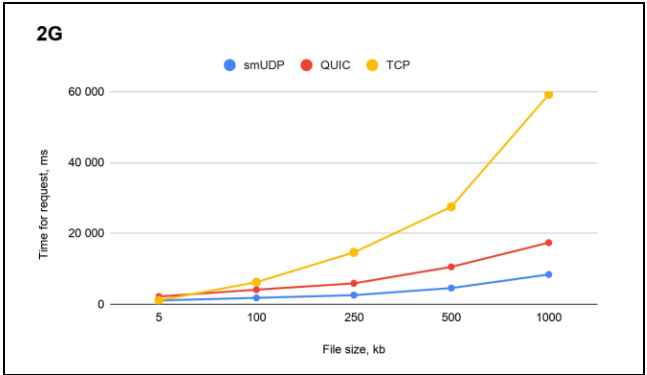

Figure 5. 2G

Additional second and third passes are spent on creating a TCP connection, while UDP and protocols based on it do not spend time on this and can send data in the first packet (see Figure 6). However, after the connection is established, the recipient continues to send a confirmation of receipt of each packet (ACK) to ensure reliable delivery.

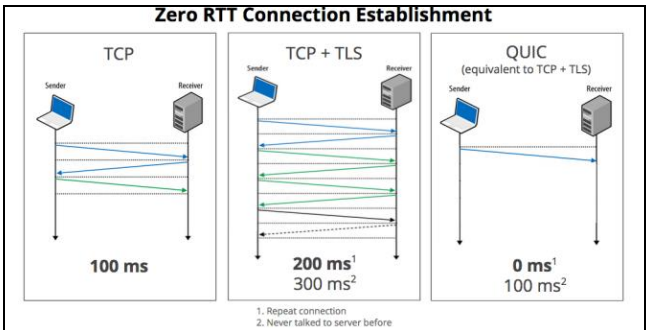

Figure 6. zeroRTT vs TCP

If a packet or ACK is lost, the sender makes a retransmission after a timeout (retransmission timeout - RTO). The RTO is calculated dynamically based on various factors, such as the expected RTT delay between the sender and receiver. At the same time, the RTT change is not expected to significantly affect the delivery speed - the growth is similar to that of UDP-based protocols, as shown in Figure 7.

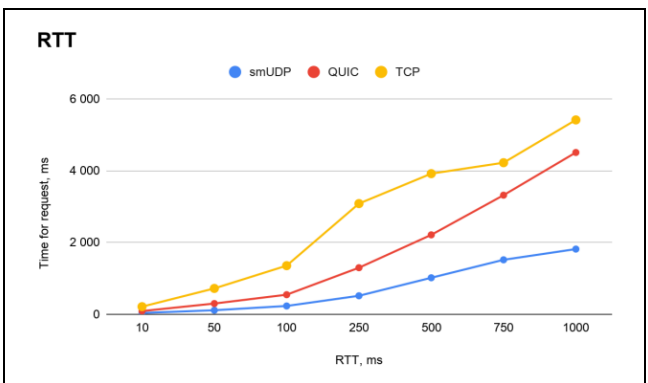

Figure 7. RTT impact comparison

However, due to the fact that some packets with ACK or the data itself may be lost due to network connection instability in TCP, the query execution time is greatly increased. This disadvantage can be clearly observed in Figure 8. In contrast to UDP-based protocols, with an increase in the percentage of lost packets, TCP also increases the time of data delivery. QUIC and smUDP demonstrate stable delivery speed, despite the presence of problems in the connection. This development is due to the fact that QUIC calls two Tail Loss Probes (TLP) before the RTO works – even when the losses are very noticeable. This is different from TCP implementations. TLP mainly forwards the last packet (or a new one, if there is one) to start fast replenishment. This is also due to the fact that TCP was originally developed as a protocol for wired network connections - which is more stable than wireless networks [10]. Wireless networks are designed differently. To deal with fluctuations in bandwidth and loss, wireless networks usually use large buffers for traffic spikes. TCP very often treats the queue as a loss due to the increased response timeout, which is why TCP is forced to retransmit packets, which leads to a full buffer and a longer query execution time.

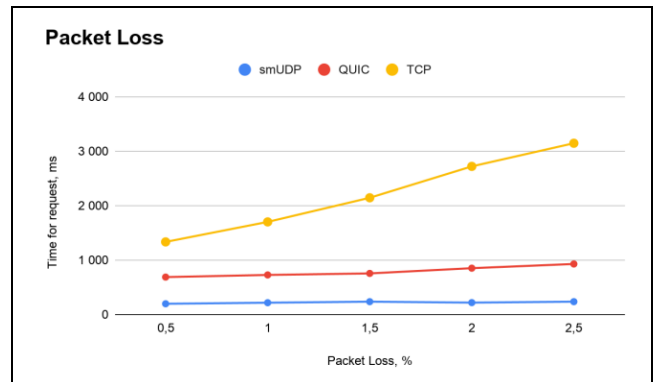

Figure 8. Packet Loss impact comparison

When studying the stability of protocols to change bandwidth, we can note that, in general, this parameter does not significantly affect the transfer speed. This is due to the fact that the TCP and QUIC protocols contain an implementation of Congestion Control that limits the channel load, as shown in Figure 9. This saves the network from being overloaded. This is because, if the network is overloaded, it is quite likely that data is sent, some packets do not reach the destination, then more data is sent, and all this data is lost again. Congestion Control is responsible for limiting the output of data in certain portions.

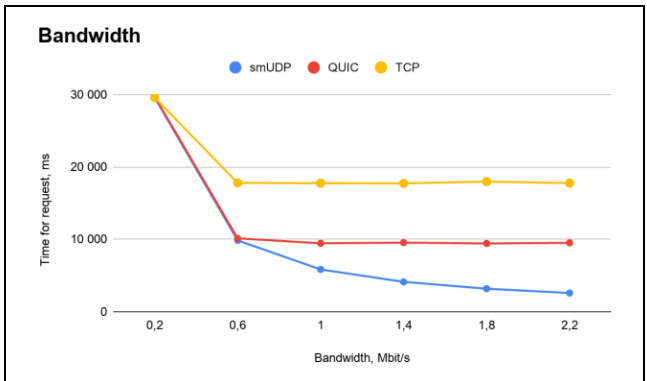

Figure 9. Bandwidth impact comparison

## V. CONCLUSION

Based on the research in this paper, we can conclude that the TCP data transfer protocol is poorly optimized for wireless networks, which are currently very widely used and are rapidly spreading around the world, even in the most remote corners of it. Poor performance is associated with unstable wireless network connections. As a result, the data reaches the end customer over a longer period of time. However, the protocol provides guaranteed data delivery to the client.

At the same time, UDP-based protocols demonstrate good performance in fast data delivery to the client due to the fact that new solutions are implemented in new protocols separately by each protocol developer, and to use the protocol, it is enough to update the versions on the server and client. However, as for data integrity, this task falls on the shoulders of the one who will use this protocol for their own purposes.

It is also worth noting that there are already ready-made implementations of UDP-based protocols that are rapidly gaining popularity, such as QUIC. It is worth noting that QUIC is already used on 4.3% of all websites [11].

Thus, if the content on a service or resource is mostly consumed via wireless mobile networks, it is recommended to use a UDP-based protocol with its own implementation of the necessary mechanisms. However, if not enough resources are available to implement one's own protocol, one can use ready-made solutions. TCP can be used as a backup connection.